# Intriguing complex magnetism of Co in RECoAsO (RE=La, Nd and Sm)

Anand Pal[1,2], M. Tropeano[2], Mushahid Hussain[3], Hari Kishan[1] and V.P.S. Awana[1,*]

[1] Quantum Phenomenon and Applications (QPA) Division, National physical Laboratory (CSIR) Dr. K.S. Krishnan Marg, New Delhi-110012, India

[2]CNR-SPIN, Universita de Genova, via Dodecaneso 33, 1-16146 Genova, Italy

[3] Department of Physics, Jamia Millia Islamia University, New Delhi-110025, India

We synthesized bulk polycrystalline samples of RECoAsO (RE=La, Nd and Sm) by solid state reaction route in an evacuated sealed quartz tube. All these compounds are crystallized in a tetragonal structure with space group *P*4/*nmm*. The Co, in these compounds is in itinerant ferromagnetic state with its paramagnetic moment above 1.5$\mu_B$ and the same orders ferromagnetically (*FM*) with small saturation moment of around 0.20$\mu_B$ below say 80K. This bulk intrinsic magnetism of Co changes dramatically when nonmagnetic La is changed by magnetic Sm and Nd. Although the itinerant ferromagnetism occurs below 80-100K with small saturation moment, typical anti-ferromagnetic (*AFM*) transitions ($T_{N1}$, $T_{N2}$) are observed at 57K and 45K for Sm and at 69K and 14K for Nd. The transition of Co spins from *FM* to *AFM*, for magnetic Sm and Nd in RECoAsO is both field and temperature dependent. For applied fields below 100Oe, both $T_{N1}$ and $T_{N2}$ are seen, with intermediate fields below 1-2kOe only $T_{N1}$ and above say 5kOe the *AFM* transition is not observed. This is evidenced in isothermal magnetization (*MH*) plots as well. It is clear that Sm/Nd magnetic moments interact with the ordered Co spins in adjacent layer and thus transforms the *FM* ordering to *AFM*. All the studied compounds are metallic in nature, and their magneto-transport *R(T)H* follows the temperature and field dependent *FM-AFM* transition of ordered Co spins.

Introduction

The discovery of superconductivity in doped oxy-pnictides; REFePO [1] and REFeAsO [2] had taken the condensed matter physics community by surprise. Interestingly these oxy-pnictides were known decades before, but in un-doped non superconducting form. Synthesis and structural properties of these compounds are reviewed recently due to renewed interest in them [3, 4]. Superconductivity basically resides in carrier doped Fe-As/Fe-P layer and hence role of 3d metal, Fe is important. Soon after, the search for superconductivity started in similar compounds with complete substitution of Fe with other 3d metals such as Co [5-12]. Though the iso-structural RECoAsO are not superconducting, their magnetic properties, including itinerant ferromagnetism [5-8] and interacting $RE^{4f}$ and $Co^{3d}$ moments [9-13] have been in discussion recently. Here we present a comparative compact study of the magnetization, magneto-transport and heat capacity of both non magnetic (LaCoAsO) and magnetic (Sm/NdCoAsO) oxy-pnictides. Both the magnetization and magneto-transport properties change drastically when non magnetic La is replaced by magnetic Sm and Nd.

Experimental

Polycrystalline samples of compositions LaCoAsO, NdCoAsO and SmCoAsO are synthesized by single step solid-state reaction method via vacuum encapsulation technique [11]. High purity (~99.9%) rare earths (La, Sm, Nd) As, $Co_3O_4$ and Co powders are ground thoroughly using mortar and pestle. The mixed powders were palletized and vacuum-sealed ($10^{-4}$ Torr) in a quartz tube. These sealed quartz ampoules were placed in box furnace and heat treated at 550°C for 12 hours, 850°C for 12 hours and then finally at 1150°C for 33 hours in continuum. The X-ray diffraction patterns are taken on Rigaku X-ray diffractometer with Cu $K_\alpha$ radiation. All physical property measurements, including magnetization, magneto-transport and heat capacity are carried out on Quantum Design PPMS (Physical property measurement system) with fields up to 14 Tesla.

Results and Discussion

The LaCoAsO, NdCoAsO and SmCoAsO compounds being used in present study are all crystallized in P4/nmm space group with lattice parameters *a* and *c* as 4.086(3)Å , 8.358(2)Å for La, 3.965(1)Å, 8.277(4)Å for Nd and 3.957(3)Å, 8.242(2)Å for Sm. These are in agreement with earlier reports [5-13]. The lattice parameters clearly follow the ionic size variation of trivalent La, Nd, and Sm.

Figure 1(a), (b) and (c) represent the magnetization data of LaCoAsO, NdCoAsO and SmCoAsO compounds respectively. Qualitatively speaking, the LaCoAsO exhibits simple paramagnetic (*PM*) - ferromagnetic (*FM*) transition at around 80K, on the other hand the NdCoAsO and SmCoAsO undergo successive *PM-FM-AFM* transitions. The isothermal magnetization (*MH*) plots for LaCoAsO, NdCoAsO and SmCoAsO compounds are given in insets of Figures (a), (b) and (c) respectively. In *PM* state Co carries a magnetic moment above 1.5$\mu_B$/Co with a *FM* saturation moment of around 0.20$\mu_B$/Co. The Co is clearly in itinerant ferromagnetic state. This is in general agreement with earlier reports on these systems [5-13]. More specifically, figure 1(a) depicts the *MT* plots for LaCoAsO at various fields of 10Oe, 500Oe and 10kOe. The compound orders *FM* with Curie temperature ($T_c$) below 70K. With increase in field the $T_c$ increases which is usual for a normal *FM* transition. The inset exhibits the *MH* plots at 5, 50, 100 and 200K. At 200K and 100K the compound is paramagnetic with linear *MH* plots and at lower temperatures (5, 50K), remains ferromagnetic with nearly no coercive field ($H_c$). The *MT* plots for NdCoAsO are depicted in Fig. 1(b) at 50Oe, 500Oe, 5kOe and 50kOe. NdCoAsO is *PM* above 70K, *FM* between 70K down to 20K and *AFM* below 20K. This situation is same for low applied fields up to 500Oe, and changes slightly at higher field of 5kOe. At 50kOe, the successive *FM* and *AFM* transitions are not clearly evident, and the resultant behavior is more like a *PM*, though not exactly. This happens due to canting of ordered moments under higher fields. The *MH* plots of NdCoAsO are shown in inset of Fig. 1(b), which are linear in *PM* (80K), slightly off axis in *AFM* (2.5K) states and are *FM* like between 12-50K with small opening of the loops. Unlike as in LaCoAsO, the *FM* plots are not saturated because of the contribution from trivalent magnetic Nd ions along with the ordered Co spins to the total moment of the studied NdCoAsO compound.

The *MT* plots of SmCoAsO in applied fields of 10Oe, 100Oe, 1kOe and 50kOe are given in Fig. 1(c). The compound undergoes successive *PM-FM-AFM* transitions with lowering the temperature, similar to that as in case of NdCoAsO. However there is a difference, the *FM* regime in NdCoAsO is quite wide (70K-20K) under an applied field as high as 500Oe. On the other hand in SmCoAsO, the *FM* regime is narrow, not more than 10K, even at low field of 10Oe [11]. It is clear from Fig. 1(c) that *FM* to *AFM* transformation in SmCoAsO is peak like even at 100Oe. The *MH* plots of SmCoAsO are shown in inset of Fig. 1(c), which are linear in *PM* (80K, 100K) and *AFM* states (2.5K, 5K). In intermediate regime, i.e., 8, 10, 20, 30, 40 and 50K the *MH* are *FM* like, but with some interesting differences. For example thought at 30, 40 and 50K the *MH* plots are purely *FM* like; at 10 and 20K the meta-magnetic like shallow steps are seen. Further at 8K, the compound seems to be transforming to *AFM* state with decreased *FM* saturation moment. It seems, as if the compound is passing through a competing *FM-AFM* transformation. Also small hysteresis is observed in *MH* at 20K and 20kOe field at the meta-magnetic transition shoulder.

Summing up, the magnetization results of RECoAsO (RE = La, Nd and Sm), though all the three are itinerant ferromagnets [5-13], the Nd and Sm undergoes successive *FM-AFM* transformation at low temperatures. The nature of transformation (*FM-AFM*) appears qualitatively same for both Nd and Sm, yet some intrinsic differences between two have been observed.

The resistivity versus temperature ($\rho(T)$) plots for LaCoAsO at various applied fields of 0, 5, 10, 50 and 140kOe are depicted in Fig. 2(a). The compound is metallic in the studied temperature range from 2.5K to 300K. Reasonable magneto-resistance (*MR*) appears below Currie temperature ($T_c$) and is maximum at 60K. The *MR* of LaCoAsO at different temperatures is shown in inset of Fig. 2(a). The *MR*% is maximum to ~ -12% at 50K in 100kOe field with a cusp-shape. *MR*% is decreased at both higher (100K, 200K) and lower temperatures (20, 10, 5 and 2.5K). At 100K and 100kOe field the maximum *MR* is -5% and -4% at 100K and 20K respectively. *MR* is negligible in pure *PM* state (200K) and also in saturated *FM* phase at low temperatures (10, 5 and 2.5K). This is usual as the MR is maximum only in the magnetic phase separated region when *FM* conversion is taking place and is negligible in pure *PM* (200K) and saturated *FM* (10, 5 and 2.5K) phases [14]

The $\rho(T)$ plots for NdCoAsO at 0, 5, 10, 50 and 140kOe are depicted in Fig. 2(b). The compound shows similar metallic behavior to that seen for LaCoAsO with reasonable *MR* below 100K. The interesting difference is that the $\rho(T)$ of NdCoAsO shows a step-like up-turn at around 18K in zero field and shifts to lower temperature (18K to 4K) with an applied field of 140kOe. This Step like transition is ascribed to the ordering of Nd$^{4f}$ moments being hybridized with Co$^{3d}$ moments [9]. Inset of Fig. 2(b) reveals the *MR* at different temperatures. The *MR* is negligible at 200K i.e., in *PM* state, and increases to -6% at 100K and 50K under applied fields of 100kOe. At 20K the *MR* is decreased and becomes -2% under 100kOe field. This is similar to that as observed for LaCoAsO, where the −*MR*% is maximum in the *PM-FM* transformation region and least in both *PM* and saturated *FM* states. The situation is very interesting below 20K, i.e., where Nd$^{4f}$ spins order *AFM* [9]. At 10K the −*MR*% is like a delta function, i.e., maximum *MR* is reached in relatively smaller field of 10kOe to -10% and remains nearly invariant till 100kOe. At 5K, the *MR* is reached to -18% at 30kOe and later remains constant. The same situation occurs at 2.5K, at higher field of 80kOe maximum *MR* is nearly -18%. The unusual increase of *MR* at lower temperatures is due to the ordering of Nd$^{4f}$ moments and their interplay with the *FM* ordered Co$^{3d}$ spins. This is unlike LaCoAsO, where non magnetic La does not influence the ordered Co spins.

Fig. 2 (c) exhibits the $\rho(T)$ plots for SmCoAsO at various applied fields of 0, 5, 10, 50 and 140kOe. The SmCoAsO is metallic in the studied temperature range of 2.5K to 300K and no step like transition is observed in zero field, unlike NdCoAsO. Interestingly the same $\rho(T)$ step like upward transition starts appearing below 45K under applied field, see Fig. (2c). The step like transition temperatures decreases to 30, 15, and below 10K for applied fields of 5, 50 and 140kOe. Because the step-like transition is of same nature for NdCoAsO at zero field and under field for SmCoAsO; one may infer that Sm$^{4f}$ spins comes to similar ordered state under field as that of Nd in zero field. Magnetic structure data of SmCoAsO is yet warranted and that could only clarify the situation. The isothermal *MR* data of SmCoAsO up to 100kOe at various temperatures of 2.5, 5, 10, 20, 50, 100 and 200K is shown in inset of Fig. 2(c). The *MR* is minimum at both *PM* (200K) and low temperature *AFM* (2.5K, 5K) states. The SmCoAsO undergoes successive *PM-FM-AFM* transitions. The *MR* is maximum during the transformation regime temperatures of 100, 50, 20 and 10K. At 100K the SmCoAsO is transforming from *PM* to *FM* state and at below 50K the competition is between *FM-AFM*. Hence qualitatively the maximum isothermal *MR* at fixed temperatures can be understood in terms of *PM-FM-AFM* magnetic transitions. Also evident in inset of Fig. 2(c) are the shoulder hysteresis during increase/decrease of field at 10K. This reminds the shoulder hysteresis seen in isothermal magnetization (*MH*) of SmCoAsO, see inset Fig. 1(c).

The heat capacity ($C_P(T)$) of LaCoAsO, NdCoAsO and SmCoAsO are depicted in Fig. 3. The $C_P(T)$ plots are smooth and as such various Co magnetic orderings are not seen. The expected $T_N{}^{Sm}$ (*AFM*) is seen as a peak in $C_P(T)$ at 5.4K, and for Nd below 2K, in SmCoAsO and NdCoAsO respectively. To elucidate on complex Co spins ordering in these samples the $C_P/T$ vs *T* plots are plotted in inset of Fig. (3). The plots slope changes first near 80K, which roughly coincides with the *FM* ordering of Co spins. With further lowering of temperature the $C_P/T$ vs T plots slope change at around 20K and 15K with a shallow broad minimum respectively for SmCoAsO and NdCoAsO. These temperatures roughly coincide with the complex *AFM* ordering of Sm$^{4f}$-Co$^{3d}$ and Nd$^{4f}$-Co$^{3d}$ interplayed matrix. It seems the Sm/Nd-CoAsO undergo three magnetic transitions i.e., $T_c^{Co}$ (~80K), the $Sm^{4f}$-$Co^{3d}$ and $Nd^{4f}$-$Co^{3d}$ interplayed *AFM* below 20K and finally $Sm^{3+}$ and $Nd^{3+}$ spins individual *AFM* at 5.4K and below 2K respectively. On the other hand in LaCoAsO, the only ordering seen is for Co spins with $T_c^{Co}$ at around 80K.

Summarily, unlike LaCoAsO, NdCoAsO and SmCoAsO undergo complex magnetic transformations from *FM* to *AFM* due to interplay of $Nd^{4f}$/$Sm^{4f}$ with $Co^{3d}$ moments. Unlike NdCoAsO [9], detailed magnetic structure of SmCoAsO is warranted.

Authors would like to thank their Director Prof. R.C. Budhani for his keen interest and encouragement for the study. This work is partly supported by Indo-Italy *CSIR-CNR* project. Mr. Ankur Rastogi from IIT(K) is acknowledged for careful reading of the manuscript. Anand Pal would like to thank CSIR for an award senior research fellowship.

References

1. Y. Kamihara, H. Harimatsu, M. Hirano, R. Kawamura, H. Yanagi, T. Kamiya and H. Hosono, J. Am. Chem. Soc. **128**, 112 (2006).
2. Y. Kamihara, T. Watanabe, M. Hirano, and H. Hosono, J. Am. Chem. Soc. **130**, 3296 (2008).
3. R. Pottgen and D. Johrendt, Z Naturforsch **63b**, 1135 (2008).
4. T.C. Ozawa and S.M. Kauzlaric, Sci Techno Adv Mater **9**, 033003 (2008)
5. A.S. Sefat, A. Huq, M.A. McGuire, R. Jin, B.C. Sales, D. Mandrus, L.M.D. Cranswick, P.W. Stephens, and K.H. Stone, Phys. Rev. B **78**, 104505 (2008).
6. H. Yanagi, R. Kawamura, T. Kamiya, Y. Kamihara, M. Hirano, T. Nakamura, H. Osawa, and H. Hosono, Phys. Rev. B **77**, 224431 (2008).
7. M. Majumdar, K. Ghoshray, A. Ghoshray, B. Bandyopadhyay, B. Pahari and S. Banerjee, Phys. Rev. B **80**, 212402 (2009).
8. H. Ohta and K. Yoshimura, Phys. Rev. B **79**, 184407 (2009).
9. M. A. McGuire, D. J. Gout, V. O. Garlea, A. S. Sefat, B.C. Sales, and D. Mandrus Phys. Rev. B **81**, 104405 (2010).
10. A. Marcinkova, D. A. M. Grist, I. Margiolaki, T. C. Hansen, S. Margadonna, and Jan-Willem G. Bos, Phys. Rev. B **81**, 064511 (2010).

11. V.P.S. Awana, I. Nowik, Anand Pal, K. Yamaura, E. Takayama-Muromachi, and I. Felner, Phys. Rev. B **81**, 212501 (2010).
12. H. Ohta, C. Michioka, A. Matsuo, K. Kindo, and K. Yoshimura, Phys. Rev. B **82**, 054421 (2010).
13. C. Krellner, U. Burkhardt, and C. Geibel, Physica B **404**, 3206 (2009).
14. M. Uhera, S. Mori, C.H. Chen, S.-W. Cheong, Nature (London) **399**, 560 (1999).

Figure Captions

Figure 1: (a), (b) and (c) shows magnetization ($M$) versus temperature ($T$) for LaCoAsO, NdCoAsO and SmCoAsO at different fields respectively. $M(H)$ of these compounds are shown in their respective insets.

Figure 2: (a), (b) and (c) shows resistivity ($\rho$) versus temperature ($T$) for LaCoAsO, NdCoAsO and SmCoAsO at different fields respectively from 300 down to 2.5K. $MR$ of these compounds at various temperatures are shown in their respective insets.

Figure 3: Heat capacity ($C_P$) versus temperature ($T$) for LaCoAsO, NdCoAsO and SmCoAsO at from 200 down to 2K. Inset shows the $C_P/T$ versus $T$ plots.

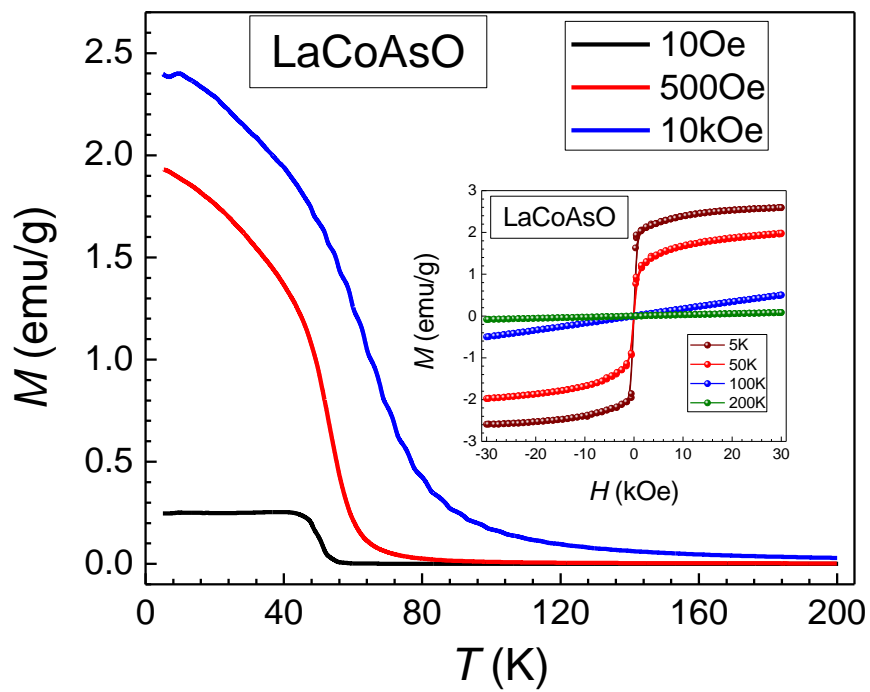

Fig. 1(a)

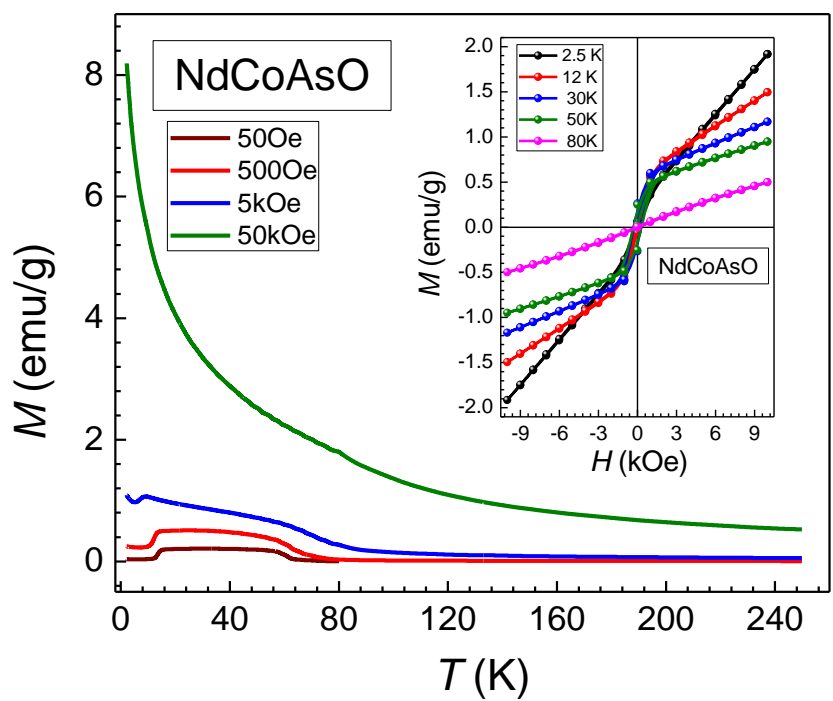

Fig. 1(b)

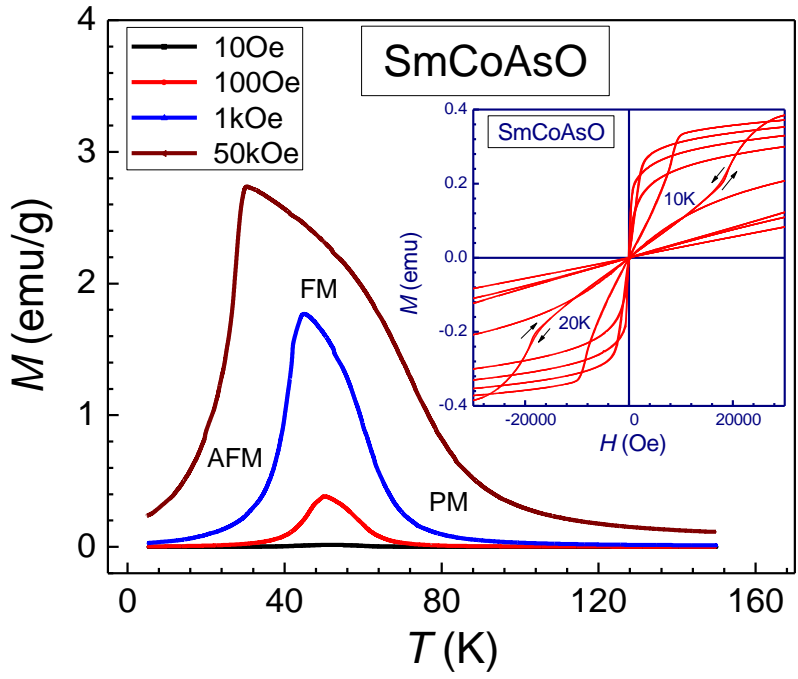

Fig. 1(c)

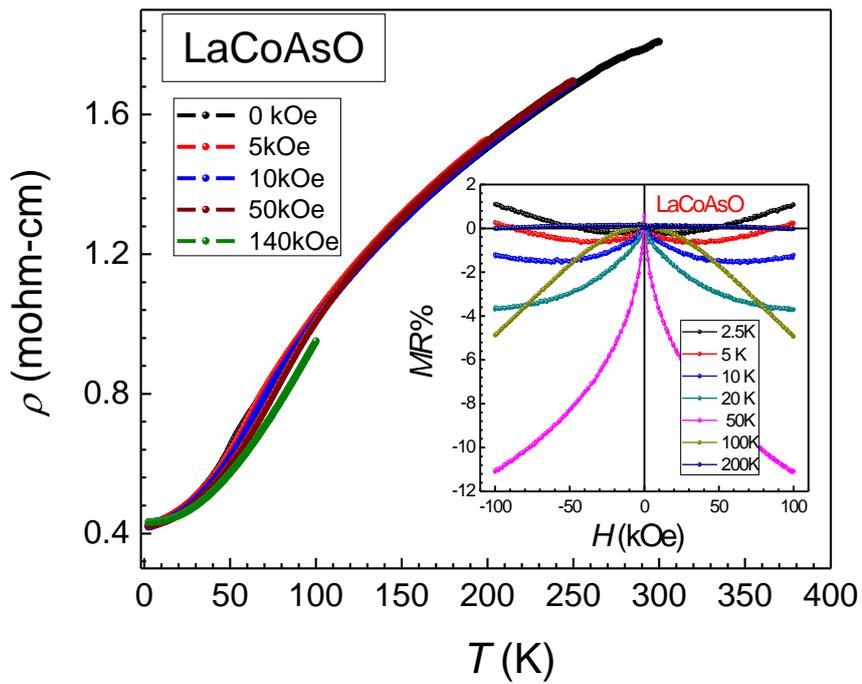

Fig. 2(a)

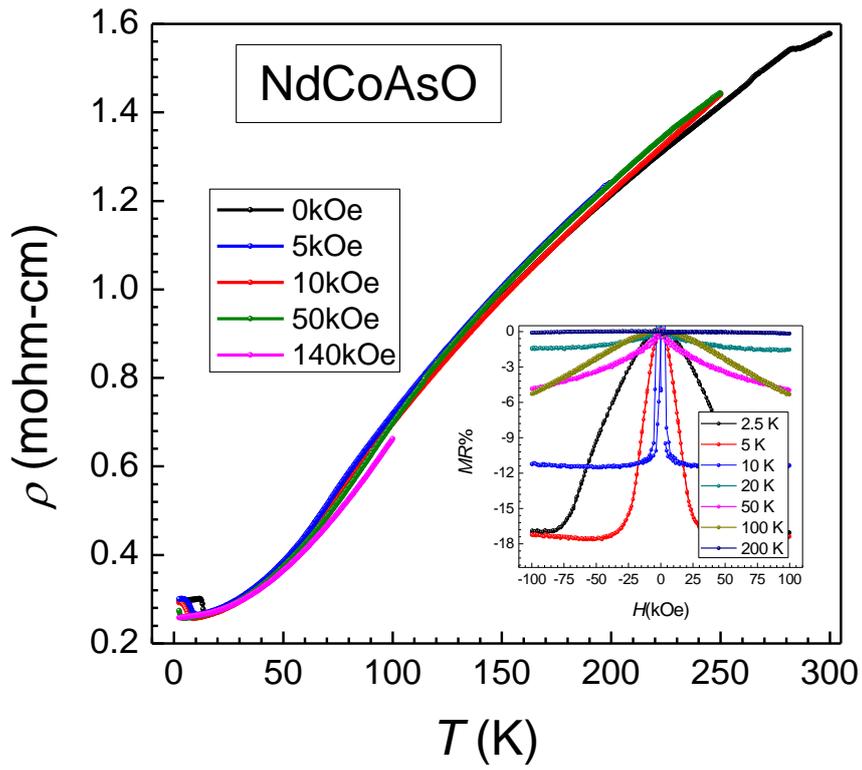

Fig. 2(b)

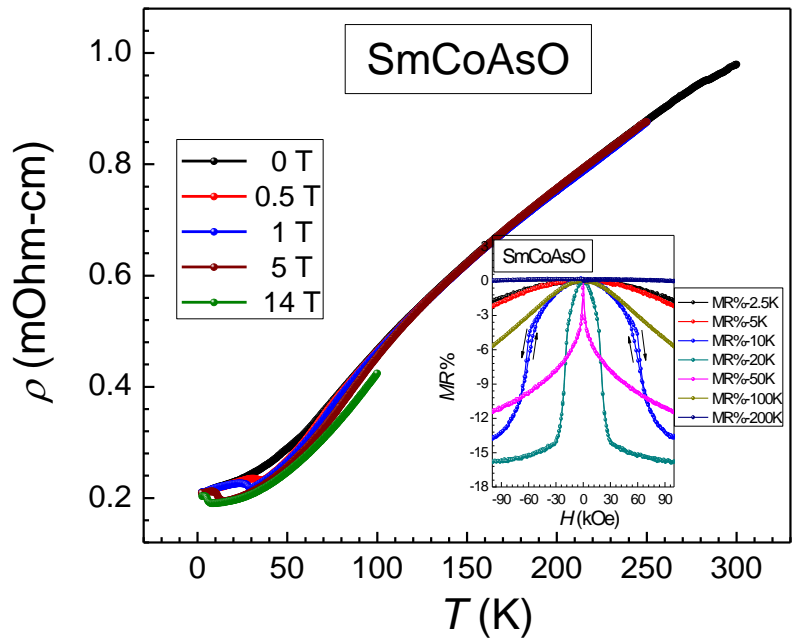

Fig. 2(c)

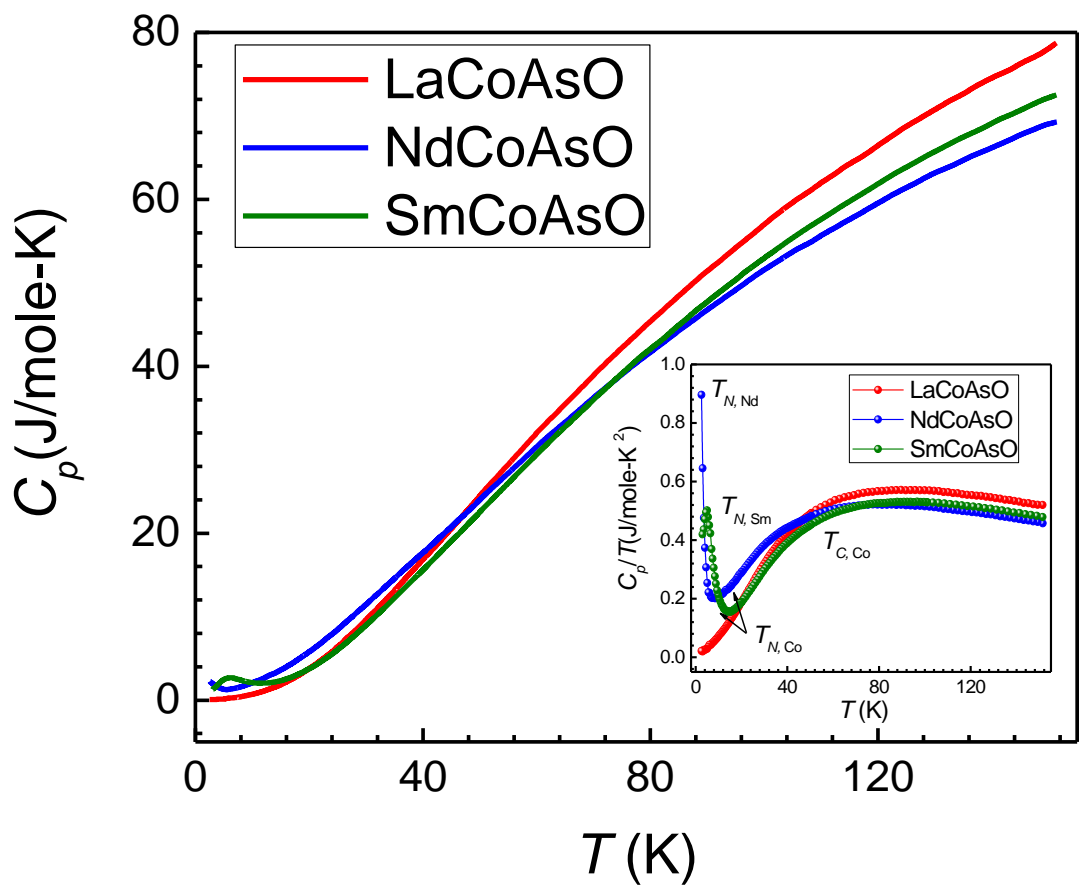

Fig. 3